\documentclass[aip,apl,reprint,twocolumn,superscriptaddress]{revtex4-1}

\usepackage{amsmath}
\usepackage{graphicx}

\begin{document}

 \title{Dispersionless propagation of electron  wavepackets in single-walled carbon nanotubes}

 \author{Roberto Rosati}
\affiliation{
Department of Applied Science and Technology, Politecnico di Torino \\
C.so Duca degli Abruzzi 24, 10129 Torino, Italy
}

\author{Fabrizio Dolcini}

\affiliation{
Department of Applied Science and Technology, Politecnico di Torino \\
C.so Duca degli Abruzzi 24, 10129 Torino, Italy
}

\affiliation{CNR-SPIN, Monte S.Angelo - via Cinthia, I-80126 Napoli, Italy}
\author{Fausto Rossi}
\affiliation{
Department of Applied Science and Technology, Politecnico di Torino \\
C.so Duca degli Abruzzi 24, 10129 Torino, Italy
}

\date{\today}

\begin{abstract}
We investigate the propagation of electron wavepackets in single-walled carbon nanotubes via a Lindblad-based density-matrix approach that enables us to account  for both  dissipation and decoherence effects induced by various phonon modes. We show that, while in semiconducting nanotubes the  wavepacket experiences the typical dispersion  of conventional materials, in metallic nanotubes its shape remains essentially unaltered, even in the presence of the electron-phonon coupling, up to micron distances at room temperature.
\end{abstract}



\maketitle

One of the most fascinating challenges in today's nanotechnology is to  control the dynamics of   electron waves with an accuracy comparable to the one available in optics.\cite{Kok07}
To this purpose, it is crucial to generate sequences of wavepackets that propagate coherently without overlapping to each other. Although single-electron pumps have been implemented,\cite{Fève07,McNeil11,Pekola08} electron waves exhibit a major limitation as compared to their electromagnetic counterpart: While the photon velocity  is independent of   its wavevector,  the electron group velocity  in conventional materials (with a parabolic-band structure) depends on it, causing the well known intrinsic dispersion of the electron wavepacket, even in the absence of scattering processes. Thus, despite various proposals,\cite{Bertoni00,Jefferson06,Fève08,Pekola13} the realization of   single-electron wavepackets of controllable shape and phase that propagate ballistically in low-dimensional conductors  still remains an open problem.

In metallic single-walled carbon nanotubes (SWNTs), in graphene  and in the recently discovered topological insulators, however, electrons behave as massless relativistic fermions and, just like photons, exhibit a linear spectrum, with the Fermi velocity $v_F \sim 10^6\,{\rm m/s}$ replacing the speed of light~$c$. This property makes such materials ideal candidates for an electronic alternative to photon-based quantum information processing.
In graphene, for instance, electron supercollimation has been predicted to occur when an external static and long-range disorder
is suitably applied.\cite{Park08,Choi14} SWNTs, nowadays synthesized with high accuracy\cite{Charlier07,Maruyama02,Arnold06} and used as versatile building blocks of high-performance electronic devices,\cite{Javey03,Kang07,Nougaret09} are even more promising.

Although in principle an electron wavepacket can propagate along a metallic  SWNT  maintaining its initial shape, any implementation in realistic devices requires to determine   whether and to what extent such property is affected by scattering processes. The high level of purity reached in carbon-nanotube synthesis\cite{Maruyama02} enables one to fairly neglect scattering with vacancies or defects. Furthermore, charged-impurity potentials arising from the substrate are typically  long-ranged over the C-C  distance, and are well known to produce no backscattering, due to the Klein tunneling effect related to the two-fold sublattice degree of freedom characterizing C-based materials. At low temperatures, electron-electron scattering has been shown to lead to Coulomb-blockade\cite{Tans98} and to Luttinger-liquid behaviors.\cite{Bockrath99} However, at intermediate and room temperature, experiments indicate that the main source of scattering  arises from electron-phonon coupling,\cite{Yao00,Park04,Javey04} whose effects have attracted the interest of theorists in last years. On the one hand, models based on a  classical-like treatment of the electron-phonon coupling (as an external oscillating potential)\cite{Roche05a,Ishii07} enable one to obtain the time-dependent evolution of  single wave packets and to extract the linear conductance by a suitable average over the initial state. These approaches, however, fail in capturing the intrinsically dissipative nature of the phonon bath. On the other hand, treating electron-phonon coupling in SWNTs via Boltzmann-equation schemes\cite{Perebeinos05} does not allow one to account for electronic phase coherence.

In this Letter  we investigate the impact of electron-phonon coupling on the dynamics of an electron wavepacket propagating in SWNTs; to this end, we employ a recently developed density-matrix approach that enables us to account for  both energy dissipation and  decoherence effects. We demonstrate that, while in semiconducting SWNTs an electronic wavepacket  spreads already for a scattering-free propagation, in  metallic SWNTs the shape of the wavepacket is essentially unaltered, even accounting for electron-phonon couplings. Differently from predictions based on the conventional relaxation-time approximation (RTA), our results thus indicate that metallic SWNTs can realistically be utilized as an electron-based platform for information transfer.

The low-energy carrier Hamiltonian of a SWNT decouples into two valleys around the $\mathbf{K}$ and $\mathbf{K}^\prime$ points, described by Dirac matrices $H_\mathbf{K}=\hbar v_F \boldsymbol\sigma \cdot  \mathbf{k}$ and $H_{\mathbf{K}^\prime}=-\hbar v_F \boldsymbol\sigma^* \cdot  \mathbf{k}$, respectively, where  $\boldsymbol\sigma =(\sigma_x,\sigma_y)$ denote Pauli matrices acting on the twofold sublattice space, and  $\mathbf{k}=(k_{n,\nu}^\perp,k)$  the  carrier wavevector.\cite{Ando05} Here $k$ denotes the continuous component along the SWNT axis  ($\parallel$), whereas $k_{n,\nu}^\perp=(n+v \nu/3)/R$ is the discrete component along the circumference ($\perp$), with $n$  denoting the electron subband, $v =\pm 1$ for  the $\mathbf{K}/\mathbf{K}^\prime$ valley, $R$ the nanotube radius, and the index $\nu = 0, \pm 1$ is defined through the relation $\exp[i \mathbf{K} \cdot \mathbf{C}]=\exp[-i \mathbf{K}' \cdot \mathbf{C}]=\exp[2\pi i \nu/3]$, where $\mathbf{C}$ is the vector rolling    the graphene lattice into the SWNT.
Since typical subband energy separations are of the order of ${\rm eV}$, we shall focus on the lowest energy subband ($n=0$), whose energy spectrum is given by
\begin{equation}\label{epsilon}
\epsilon_{\alpha} = b\,\hbar v_F \sqrt{
k^2 +( \nu/3R)^2
}\ ,
\end{equation}
where $\alpha=(k, b, v,\nu)$ is the quantum-number multi-label and $b= \pm 1$  for the conduction and valence band, respectively. The related electron eigenfunctions are $\phi_\alpha(\mathbf{r})=(4\pi R L)^{-1/2} (e^{-iv \theta_\mathbf{k}/2},bv e^{+iv \theta_\mathbf{k}/2})\, e^{i \mathbf{k} \cdot \mathbf{r}}$, where $L$ denotes the nanotube length and $\theta_\mathbf{k}$ the polar angle of $\mathbf{k}$. While for $\nu \neq 0$ the energy spectrum is gapped  (semiconducting nanotube) and near $k = 0$ is parabolic-like similarly to conventional semiconductors, for $\nu = 0$ the spectrum is gapless (metallic case), and the typical massless Dirac-cone structure is recovered. All armchair and (3n,0) zigzag SWNTs are remarkable examples of the metallic case.\cite{Ando05}

The SWNT phonon spectrum near the $\mathbf{K}$ and $\mathbf{K}^\prime$ points includes\cite{Ando05,Suzuura02} i)   longitudinal(L) stretching modes, characterised by an  acoustic(A) branch $\omega_{q,LA}=v_S |q|$ with $v_S \simeq 1.9 \cdot 10^4 {\rm m/s}$  and an  optical(O) branch with $\hbar \omega_{LO}\simeq 0.2$ eV; ii) breathing(Br) modes orthogonal to the nanotube surface, with a roughly $q$-independent spectrum $\hbar \omega_{Br} \simeq 0.14 \,{\rm eV \AA }/R$;  (iii)   transverse(T) twisting modes, characterised by an acoustic  branch with $\omega_{q,TA}=v_T |q|$ with $v_T\simeq 1.5 \cdot 10^4 {\rm m/s}$ and an  optical(O) branch with $\hbar \omega_{TO}\simeq 0.2$ eV.  For each mode the  electron-phonon coupling  is described by a $2\times 2$ matrix acting on the sublattice space, which exhibits both diagonal (intra-sublattice) and off-diagonal (inter-sublattice) entries  for the LA, TA and Br modes,\cite{Suzuura02} and is purely off-diagonal for optical modes.\cite{Ishikawa06} We neglect zone-edge phonon modes that lead to inter-valley scattering because they may become important only at very high energies.

In order to account for energy dissipation and decoherence induced by the nanotube phonon bath on the otherwise phase-preserving electron dynamics, we have applied to the carbon nanotube the  formalism   introduced in Ref.~[\onlinecite{Rosati14e}] via a numerical solution of the Lindblad-based non-linear density-matrix equation (LBE)
\begin{eqnarray}
\lefteqn{\frac{d \rho_{\alpha_1\alpha_2}}{d t} =\frac{\epsilon_{\alpha_1}-\epsilon_{\alpha_2}}{\imath\hbar}
\rho_{\alpha_1\alpha_2} + } & & \nonumber \hspace{2cm}\\
& &+ \frac{1}{2} \left[\!\sum_{\alpha'\alpha'_1\alpha'_2} \left(\left(\delta_{\alpha_1\alpha'} - \rho_{\alpha_1\alpha'}\right)
\mathcal{P}^s_{\alpha'\alpha_2,\alpha'_1\alpha'_2} \rho_{\alpha'_1\alpha'_2}
  \right.\right.\label{DME}
 \\
& &-\left.\left.
\left(\delta_{\alpha'\alpha'_1} - \rho_{\alpha'\alpha'_1}\right) \mathcal{P}^{s *}_{\alpha'\alpha'_1,\alpha_1\alpha'_2} \rho_{\alpha'_2\alpha_2}\right)\ +\ {\rm H.c.}\right] \, .\hspace{1cm}\nonumber
\end{eqnarray}
Here, the first term describes the scattering-free propagation, whereas the second term
is a non-linear scattering superoperator expressed via generalized scattering rates $\mathcal{P}^s_{\alpha_1\alpha_2,\alpha'_1\alpha'_2}$,
whose explicit form is microscopically derived from  the electron-phonon Hamiltonians described above, with $s$ labelling the various phonon modes.
The fully quantum-mechanical density-matrix equation (\ref{DME}) enables us to go beyond   the conventional  Boltzmann transport equation, which is recovered  in the diagonal limit
($\rho_{\alpha_1\alpha_2} = f_{\alpha_1} \delta_{\alpha_1\alpha_2}$),\cite{Rosati14b} where the generalized scattering rates reduce to the semiclassical rates provided by the standard Fermi's golden rule,
$P^s_{\alpha\alpha'} = \mathcal{P}^s_{\alpha\alpha,\alpha'\alpha'}$.

In order to show that carbon nanotubes can be utilised as quantum-mechanical channels for the non-dispersive transmission of electronic wavepackets, we have performed simulated experiments where the shape of an initially prepared  wave packet is monitored while it evolves under the effect of the phonon bath.   Any initial state can always be written as $\rho=\rho^\circ+\delta \rho$, where $\rho^\circ$ is the homogeneous equilibrium state and   $\delta \rho$ describes a localised excitation, whose spatial shape (e.g. Gaussian) can in principle be generated experimentally via a properly tailored optical excitation. While the description of the specific optical generation is beyond the aim of the present paper, the localisation of the initial wave packet is the crucial aspect in our analysis. A simple  choice that captures this essential physical ingredient is an initial state described by an intra-valley conduction-band density matrix $\rho_{\alpha_1 \alpha_2}=\delta_{v_1,v_2} \delta_{b_1,1} \delta_{b_2,1} \rho_{k_1\, k_2}$, where
\begin{equation}\label{is}
\rho_{k_1 k_2} = \sqrt{f^\circ_{{k}_1} f^\circ_{{k}_2}}\, e^{-\ell \vert{k}_1-{k}_2\vert}\ .
\end{equation}
Here  $f^\circ_{k}$ is the Fermi-Dirac distribution of the conduction-band states ${k}$, and the parameter $\ell$ plays the role of a delocalization length. Indeed for $\ell \to \infty$ the spatially homogeneous  equilibrium state $\rho^\circ_{{k}_1{k}_2} = f^\circ_{{k}_1} \delta_{{k}_1{k}_2}$ is recovered, whereas for finite $\ell$ the excitation $\delta \rho$ consists of interstate phase coherence (intraband polarization) determined by the parameter $\ell$. In particular, in the limit $\ell \to 0$ one obtains the maximally localized wavepacket. 
Moreover, a distinguished feature of the initial condition in (\ref{is}) is the absence of nonequilibrium diagonal contributions ($\rho_{kk} = f^\circ_k$), which implies that energy dissipation and decoherence will affect the non-diagonal contributions only; this is the typical situation produced by a weak interband optical excitation.
   \\

\begin{figure}[h]
	\centering
\includegraphics[width=\columnwidth]{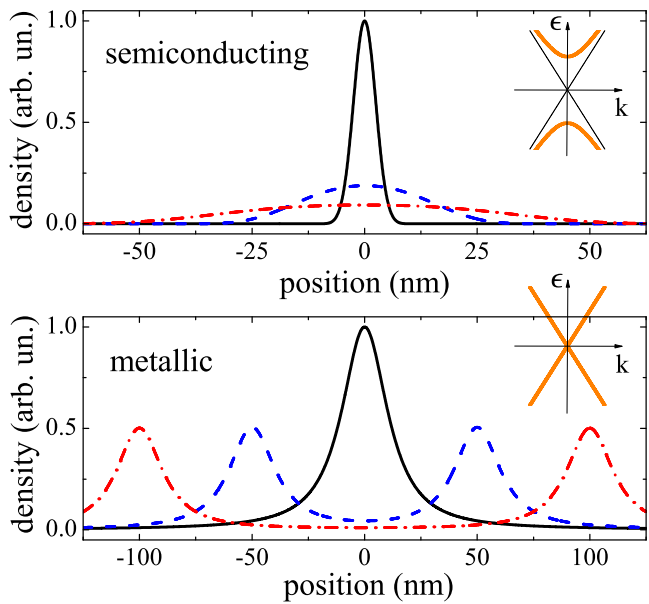}
	\caption{(Color online) Room-temperature scattering-free evolution in a semiconducting (10,0) SWNT ($\nu=1$) (upper panel) and a metallic (12,0) ($\nu=0$) SWNT (lower panel): Charge density of the maximally-localized ($\ell \to 0$) electronic wavepacket in Eq. (\ref{is}) as a function of the position along the SWNT axis  at three different times: $t=0$ fs (solid), $t=50$ fs (dashed) and $t=100$ fs (dashed-dotted). }
	\label{Fig1}
\end{figure}

We start our discussion analysing the scattering-free propagation of the initial state (\ref{is}), e.g. switching off the electron-phonon interaction in Eq. (\ref{DME}). Figure~\ref{Fig1} shows the spatial carrier density $n(r_\parallel)$ at different times for the initial state (\ref{is}), taken at room temperature and in the maximally-localized  limit ($\ell \to 0$).
The cases of semiconducting and metallic nanotube are shown in the upper and lower panels, respectively. For the  semiconducting nanotube ($\nu \neq 0$)  the dispersion relation (see Eq.~(\ref{epsilon}))  at small $k$ reduces to the parabolic spectrum of conventional semiconductor materials and gives rise to the typical classical-like diffusion process,  preventing any information transfer via electronic wavepackets. In contrast, for the case of the metallic nanotube ($\nu =0$), characterised by a linear dispersion, the initial charge peak splits into two identical and shape-preserving components which travel in opposite directions with velocity $\pm v_F$ (lower panel), i.e., $n(r_\parallel,t)=n^+(r_\parallel- v_F t)+n^-(r_\parallel+v_F t)$. This is the non-dispersive propagation scenario we are looking for.

 We thus focus on the metallic nanotube and switch on the electron-phonon coupling to address the crucial question of whether and how energy dissipation and decoherence modify such ideal dispersion-free scenario. To this end, we have performed a set of simulated experiments based on the LBE (\ref{DME}), including LA, TA, Br, LO and TO phonons. In order to expand the space scale with respect to the ideal scenario displayed in Fig.~\ref{Fig1}, here the delocalization length $\ell$ in the initial condition (\ref{is}) is chosen such to get a FWHM value of the initial peak of $0.4\mu$m.

\begin{figure}[h]
	\centering
\includegraphics[width=\columnwidth]{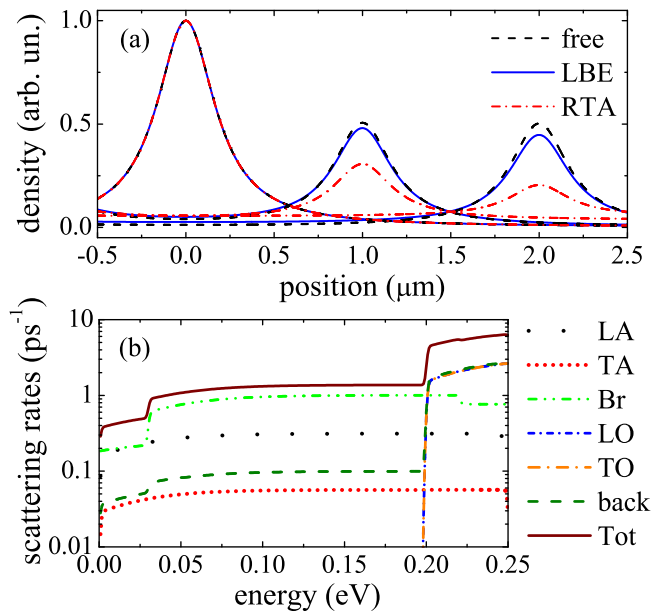}
	\caption{(Color online)
(a) Room-temperature charge-density evolution of an electronic wavepacket  in a metallic (12,0) SWNT as a function of the position $r_\parallel$ along the SWNT axis, at three different times: $t=0$ ps (left peaks), $t=1$ ps (central peaks) and $t=2$ ps (right peaks). The solid lines show the  effects of the  carrier-phonon coupling, accounted for by the LBE (\ref{DME}), compared to the scattering-free propagation (dashed lines). The RTA results (dot-dashed lines) are also plotted for comparison. Here the  wavepacket delocalization length $\ell$ in Eq. (\ref{is}) is chosen such that the FWHM initial value is 0.4 $\mu{\rm m}$.
(b) Room-temperature total scattering rates in a metallic (12,0) SWNT   as a function of the   axial energy $\epsilon^\parallel=\hbar v_F |k|$ for various electron-phonon mechanisms $s=$LA (dotted line),  TA (short-dotted line),  Br (dot-dot-dashed line),  LO (short-dot-dashed line),  TO (dot-dashed line),
as well as for their sum (solid line) and its backward component (dashed line). Note that the $s=$LO and $s=$TO lines almost coincide.
}
	\label{Fig2}
\end{figure}

Figure \ref{Fig2}(a) shows the wavepacket propagation in~a~(12,0) SWNT at room-temperature.
The electron-phonon coupling, accounted for by our LBE simulation (solid curve), does not significantly alter the shape with respect to the ideal  scattering-free  result (dashed curve), so that the transmission is essentially dispersionless up to the micrometric scale. In order to understand the origin of such shape-preserving dynamics, we first observe that in the electron-phonon coupling a natural distinction arises  between forward and backward scattering processes, where the initial and final electronic states have the same and opposite velocity sign, respectively. A semiclassical argument based on Boltzmann  theory leads to expect that in metallic SWNTs, where the group velocity can only take  two $k$-independent  values $\pm v_F$,  forward processes never contribute to the dispersion of wavepackets, which is thus due to backward processes only. 
  In our fully quantum-mechanical LBE approach the situation is less trivial, as the semiclassical transition $k \to k+q$ between purely diagonal entries of the density matrix extends to off-diagonal quantum transitions $(k_1,k_2) \to (k_1+q,k_2+q)$, so that  forward and backward processes can in principle interplay in Eq.~(\ref{DME}). In semiconducting materials such transitions can lead to  scattering nonlocality and quantum diffusion speed-up phenomena;\cite{Rosati14b} moreover, in Luttinger liquids the forward component of the electron-phonon coupling can lead to Wentsel-Bardeen  instabilities of the electron propagator.\cite{LL} 
However, for  a metallic SWNT,  a detailed investigation (not reported here) shows that, due to a strong cancellation between in- and out-scattering terms, the forward-process contribution to the scattering superoperator in (\ref{DME}) has an extremely small impact on the wavepacket propagation.    This enables us to support on firmer grounds the conclusion that in metallic SWNTs the wavepacket dispersion (solid line in Fig.~\ref{Fig2}(a)) originates mainly from backward processes. At room temperature, the latter are ascribed to breathing phonons and  to different acoustic modes, depending on the type of SWNT: for a zigzag SWNT, like the one considered in Fig.\ref{Fig2}(a), they are due to LA modes, whereas for an armchair SWNT they are due to TA modes.

Notably, a strong suppression of the wave packet would be predicted by a conventional RTA model, as shown by the dash-dotted line in Fig.~\ref{Fig2}(a). Indeed, such RTA is formulated in terms of total scattering rates $\Gamma_\alpha^s$ (from an initial state $\alpha$ to any final state) for each phonon mode $s$. The energy dependence of $\Gamma_\alpha^s$, displayed by the solid line in Fig.\ref{Fig2}(b), shows that  the most relevant scattering mechanisms are due to LA and Br modes, leading to the wrong conclusion that the typical phonon-induced wavepacket dispersion time-scale is about $2$\,ps, in accordance to the RTA result (dash-dotted line in Fig.~\ref{Fig2}(a)).
However,   the dominant contributions of LA and Br modes originate  from forward processes.   Indeed,  in the $2 \times 2$ sublattice matrix  characterising  the SWNT electron-phonon interaction for both acoustic and breathing modes, the coupling constant of the diagonal entries is known to dominate by a factor $\sim 15-25$ the off-diagonal ones.~\cite{Ando05,Suzuura02} Because backward scattering processes can only arise from  off-diagonal entries, the total backward component of the scattering rate  is at least one order of magnitude smaller than the total rate (see dashed line in Fig.~\ref{Fig2}(b)). For the reasons discussed above, namely a strong cancellation between in- and out-scattering terms (absent within the RTA model), these processes have in fact negligible impact on the diffusion process. 
We stress that also LO modes induce backward processes; however, due to their high phonon energy,\cite{Yao00} their effect is negligible.
We also notice that, at lower temperatures, the impact of electron-phonon coupling decreases and the diffusion time becomes even longer (for instance, at $77$\,K the wavepacket travels for more than one micron with a signal attenuation less than 1\%).

Finally, we emphasise that, compared to wavepacket propagation in parabolic-like systems (e.g., conventional semiconductor materials~\cite{Rosati14b} or semiconducting nanotubes (see upper panel in Fig.~\ref{Fig1})), in a metallic SWNT the impact of phonon-induced diffusion on the wavepacket propagation is very limited; indeed, while for a parabolic-like system the free-diffusion spatial-broadening displayed in the upper panel of Fig.~\ref{Fig1} is further increased due to its coupling with various dissipation channels, here the linear nature of the spectrum mostly preserves the shape of the initial peak, which is thus free to move at Fermi-velocity remaining almost unaltered despite the presence of dissipation/decoherence processes. Our results thus indicate that metallic SWNTs are a promising platform to realise quantum channels for the non-dispersive transmission of electronic wavepackets.

\medskip\par

We gratefully acknowledge funding by the Graphene@PoliTo laboratory of the Politecnico di Torino, operating within the European FET-ICT Graphene Flagship project (www.graphene-flagship.eu). F.D. also  acknowledges financial support from Italian FIRB 2012 project HybridNanoDev (Grant No.RBFR1236VV).

\end{document}